# Superconductivity in epitaxially grown LaVO$_3$/KTaO$_3$(111) heterostructures

Yuan Liu[1#], Zhongran Liu[2#], Meng Zhang[1], Yanqiu Sun[1], He Tian[2,3], and Yanwu Xie[1,4]*

[1]Interdisciplinary Center for Quantum Information, State Key Laboratory of Modern Optical Instrumentation, and Zhejiang Province Key Laboratory of Quantum Technology and Device, School of Physics, Zhejiang University, Hangzhou 310027, China

[2]Center of Electron Microscope, State Key Laboratory of Silicon Materials, School of Materials Science and Engineering, Zhejiang University, Hangzhou, 310027, China

[3]School of Physics and Microelectronics, Zhengzhou University, Zhengzhou 450052, China

[4]Collaborative Innovation Center of Advanced Microstructures, Nanjing University, Nanjing 210093, China

*To whom correspondence should be addressed. E-mail: ywxie@zju.edu.cn

#These authors contributed equally to this work.

## Abstract

Complex oxide heterointerfaces can host a rich of emergent phenomena, and epitaxial growth is usually at the heart of forming these interfaces. Recently, a strong crystalline-orientation-dependent two-dimensional superconductivity was discovered at interfaces between KTaO$_3$ single-crystal substrates and films of other oxides. Unexpectedly, rare of these oxide films was epitaxially grown. Here, we report the existence of superconductivity in epitaxially grown LaVO$_3$/KTaO$_3$(111) heterostructures, with a superconducting transition temperature of ~0.5 K. Meanwhile, no superconductivity was detected in the (001)- and (110)-orientated LaVO$_3$/KTaO$_3$ heterostructures down to 50 mK. Moreover, we find that for the LaVO$_3$/KTaO$_3$(111) interfaces to be conducting, an oxygen-deficient growth environment and a minimum LaVO$_3$ thickness of ~0.8 nm (~ 2 unit cells) are needed.

## 1 Introduction

With the development of advanced film fabrication technology, nowadays it has been routine to grow high-quality epitaxial films for many materials. This is particularly true for complex oxides[1–3], a big class of versatile materials that exhibit a wide range of exciting properties including superconductivity, ferromagnetism, ferroelectricity and many others. The ability to grow complex oxides epitaxially upon each other has led to many interfaces that host novel emergent phenomena[1–3]. Good examples include the LaAlO$_3$/SrTiO$_3$ interface that shows high-mobility electron gas[4], 2D superconductivity[5,6], and unusual magnetism[7–9], the high-temperature superconducting bilayers of copper oxides[10–12], and the PbTiO$_3$/SrTiO$_3$ superlattices with polar vortices[13].

Recently, an unusual 2D superconductivity was found at interfaces between KTaO$_3$ single-crystal substrates and other oxide films[14–22]. Unexpectedly, rare of these films is epitaxial: EuO is polycrystalline[15,16], and LaAlO$_3$[15,17–20], YAlO$_3$[21], and TiO$_x$[22] are amorphous. Therefore, the role of epitaxial growth in the KTaO$_3$ interface superconductivity is still elusive.

LaVO$_3$ is a Mott-Hubbard insulator[23,24]. Previous studies[25,26] showed that it can be grown epitaxially on KTaO$_3$(001) and forms a conducting interface. In this work, we grow epitaxial LaVO$_3$ thin films on all the (001)-, (110)-, and (111)-oriented KTaO$_3$ substrates and study their transport properties. We find that while all these LaVO$_3$/KTaO$_3$ interfaces can be conducting, only the (111)-interface is superconducting. However, the superconducting transition temperature $T_c$ is much lower than the corresponding values in the previous non-epitaxial interfaces[15–22]. In addition, we find that, for the LaVO$_3$/KTaO$_3$(111) interfaces to be conducting, an oxygen-deficient growth environment and a minimum LaVO$_3$ thickness of ~0.8 nm (~ 2 unit cells) are needed. Together with other control experiments, it suggests that electron transfer from oxygen vacancies in LaVO$_3$ film to KTaO$_3$ substrate is responsible for the interface conduction.

## 2 Experiment methods

KTaO$_3$ is a band insulator that has a cubic perovskite structure with a lattice constant of 0.3989 nm[27]. LaVO$_3$ has a GdFeO$_3$-type orthorhombic structure with lattice parameters of a = 0.5555 nm, b= 0.7849 nm, and c = 0.5553 nm at room temperature[23,24]. This structure can be regarded as a pseudocubic subcell with a lattice parameter $a_p \approx a/\sqrt{2} \approx b/2 \approx c/\sqrt{2} \approx$0.3926nm. The LaVO$_3$ thin films were grown in a high vacuum pulsed laser deposition chamber with a base pressure of ~ 4 × 10$^{-8}$ mbar. The (001)-, (110)-, and (111)-oriented KTaO$_3$ single-crystal substrates were purchased from Hefei Kejing Materials Technology Co. Ltd. A LaVO$_4$ ceramic target was ablated with a pulsed KrF excimer laser ($\lambda$ = 248 nm). Repetition rate of laser was 2 Hz and fluence at the target surface was ~3 J/cm$^2$. The growth temperature was 750 ℃. If not specified, the growth was performed at $P$(O$_2$) = 0 mbar. The film thickness was controlled by counting the growth laser pulses after calibrating the growth rate by small-angle x-ray reflectivity measurement (Fig. S2). After growth, the samples were cooled down to room temperature with 50℃/min under the growth atmosphere.

The atomic force microscopy data were taken on a Park NX10 system using noncontact mode. X-ray diffraction data were taken on a 3-kW high-resolution Rigaku Smart Lab system. Cross-sectional specimens for electron microscopy investigations were prepared by a FEI Quanta 3D FEG Focused Ion Beam. STEM images and EDS mappings were acquired using a spherical aberration-corrected microscope equipped with four Super-X EDS detectors (FEI Titan G2 80-200 Chemi STEM, 30 mrad convergence angle).

The contacts to the LaVO$_3$/KTaO$_3$ interfaces of transport measurements were made by ultrasonic bonding with Al wires. DC transport measurements were carried out in a commercial physical property measurement system (PPMS, Quantum Design) with a dilution refrigerator insert and in a commercial $^4$He cryostat with a $^3$He insert (Cryogenic Ltd.) with a standard four-probe method.

## 3 Results

### Structural characterizations

Atomic force microscopy images show that the surfaces of all the films are flat, with a root mean square roughness less than 0.4 nm (Fig. S1). From the fitting of small-angle x-ray reflectivity measurement, we can simulate that the interface and surface roughness is about 0.1 nm and 0.37 nm, respectively, indicating the high quality of the heterostructures. Figure 1a shows the x-ray diffraction (XRD) $\theta$-$2\theta$ scan of a typical LaVO$_3$/KTaO$_3$(111) sample. A clear and sharp LaVO$_3$(111) peak was observed, which indicates that the film is single-phase and epitaxially oriented. From the position of the (111) diffraction peak of the film, the c-axis lattice constant $d_{111}$ is found to be 0.226 nm ($\sqrt{3}$ $d_{111}$=0.392 nm) —is nearly the same as bulk LaVO$_3$[23]. Figure 1b shows a reciprocal space mapping (RSM) around the (132) perovskite peak. The vertical alignment of the film and substrate peaks indicates that the film is fully strained to the substrate. The crystalline quality of the LaVO$_3$/KTaO$_3$(111) sample was further examined by out-of-plane $\omega$-scan rocking curve. The full widths at half maximum (FWHM) is only about 0.09°(Fig. 1c). These observations suggest that the LaVO$_3$ film is epitaxially grown on the KTaO$_3$(111) substrate. Similar $\theta$-$2\theta$ scans and RSM images were found for the LaVO$_3$/KTaO$_3$(001) and LaVO$_3$/KTaO$_3$(110) samples (Figs. S3 and S4), indicating that the growths were epitaxial in all of them.

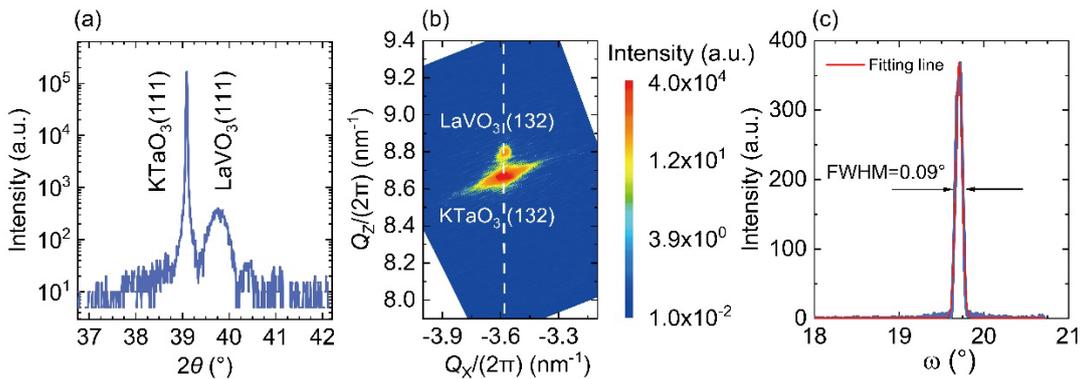

**Fig. 1** X-ray diffraction of a LaVO$_3$(17 nm)/KTaO$_3$(111) heterostructure. **a** Out-of-plane $\theta$-$2\theta$ XRD pattern. **b** Reciprocal space mapping around the (132) reflection. LaVO$_3$ and KTaO$_3$ Bragg's peaks aligned along the vertical dashed line. **c** Rocking curve taken of the LaVO$_3$ film.

The epitaxial growth of LaVO$_3$ on KTaO$_3$ is further evidenced by scanning transmission electron microscopy (STEM) measurements. In Figs. 2a and 2b we present STEM high-angle annular dark-field (HAADF) images taken from a LaVO$_3$(17 nm)/KTaO$_3$(111) sample. One can see that the LaVO$_3$ film is coherently and epitaxially grown on the KTaO$_3$(111) substrate. Note that the smeared brightness (~1-2 nm wide) along the interface is caused by interface diffusion. In our STEM-HAADF images the visible bright atomic spots are from La atoms in LaVO$_3$ and Ta atoms in KTaO$_3$. The interface diffusions of La-K and V-Ta add extra brightness in KTaO$_3$ and LaVO$_3$ sides, respectively, which result in the smeared brightness. Energy-dispersive x-ray spectroscopy (EDS) elemental mappings (Figs. 2c-2f) show that the interface diffusion is moderate, and the depth of the diffused region in each side is limited to within 1 nm from the interface. This value is much smaller than the thickness of the interface superconducting (conducting) channel[14,18,21].

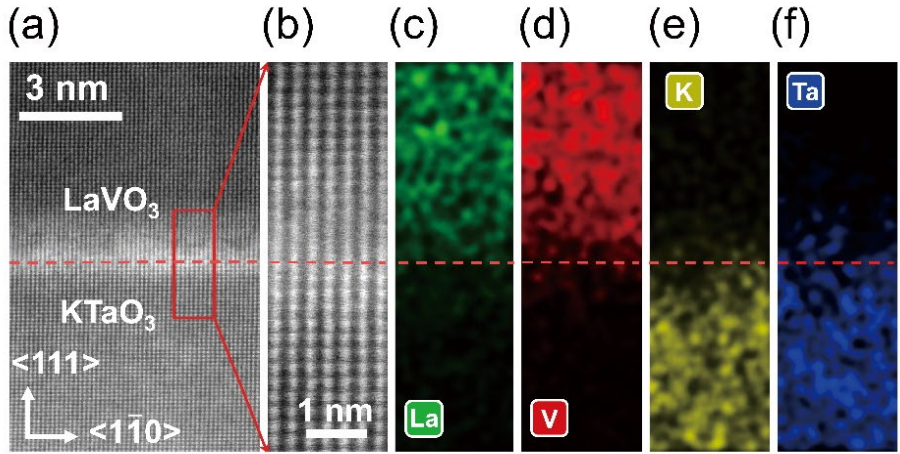

**Fig. 2** STEM characterization of a LaVO$_3$(17 nm)/KTaO$_3$(111) heterostructure. **a,b** HAADF-STEM images show that the LaVO$_3$ film is epitaxially grown on KTaO$_3$. **c-f** EDS elemental mappings of the same sample. A moderate interface diffusion is observed but the depth of the diffused region is within 1 nm in each side.

**Transport properties**

Figure 3a shows the temperature dependence of sheet resistance $R_{sheet}$ for LaVO$_3$/KTaO$_3$ heterostructures. An overall metallic conduction is observed on all of them. At a given temperature, $R_{sheet}$ decreases in a sequence following (111), (110) and (001), which is similar to that observed in previous amorphous-LaAlO$_3$/KTaO$_3$ heterostructures[17] and ionic-liquid-gated KTaO$_3$ surfaces[28]. Control experiments on LaVO$_3$ films grown on insulating (La$_{0.3}$Sr$_{0.7}$)(Al$_{0.65}$Ta$_{0.35}$)O$_3$ and NdGaO$_3$ substrates (see Figs. S5 and S6) demonstrate that the LaVO$_3$ films themselves are insulating, with $R_{sheet}$ values several orders of magnitude higher than that of the present LaVO$_3$/KTaO$_3$ heterostructures. Therefore, the observed conduction can be solely ascribed to the conducting LaVO$_3$/KTaO$_3$ interfaces.

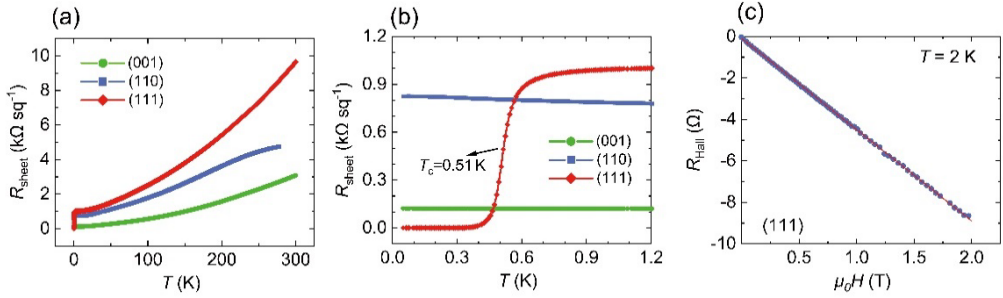

**Fig. 3** Transport properties of epitaxial LaVO$_3$(17 nm)/KTaO$_3$ heterostructures. **a** Temperature-dependent $R_{sheet}(T)$ curves from 300 K to 50 mK. **b** An enlarged view of the low-temperature data shown in **a**. **c** Hall resistance $R_{Hall}$ as a function of out-of-plane magnetic field for the (111) sample measured at 2 K.

As shown in Fig. 3b, a superconducting state with a mid-point $T_c$ of 0.51 K is observed in the $R_{sheet}(T)$ curve of the (111) sample. In contrast, no superconducting transition is detected in the (001) and (110) samples down to 50 mK. The normal-state magnetic-field-dependent Hall resistance measured at 2 K (Fig. 3c) on the (111) sample shows that the charge carriers are electrons, and the carrier density $n_{sheet}$ and Hall mobility $\mu_{Hall}$ are ~1.4 × 10$^{14}$ cm$^{-2}$ and ~42 cm$^2$V$^{-1}$ s$^{-1}$, respectively. Similarly, the measured $n_{sheet}$ and $\mu_{Hall}$ for the (001) and (110) samples are ~1.6 × 10$^{14}$ cm$^{-2}$ and ~340 cm$^2$V$^{-1}$ s$^{-1}$, and ~1.1 × 10$^{14}$ cm$^{-2}$ and ~169 cm$^2$V$^{-1}$ s$^{-1}$, respectively (not shown).

**Magnetoresistance**
The superconducting state of LaVO$_3$/KTaO$_3$(111) is further studied by magnetoresistance measurements. Figures 4a and 4b display temperature-dependent $R_{sheet}(T)$ values for magnetic fields applied perpendicular and parallel to the interface, respectively. The $T_c$ can be completely suppressed by application of a perpendicular (or parallel) field $\mu_0H$ = 0.2 T (or 2 T) (here $\mu_0$ is the vacuum permeability). Figure 4c and 4d display the temperature-dependent upper critical field $\mu_0H_{c2}$, derived from the $R_{sheet}(T)$ curves shown in Figs. 4a and 4b. The temperature dependences of both $\mu_0H_{c2\perp}$ and $\mu_0H_{c2//}$ show a linear behavior, which is different from 2D Ginzburg-Landau model and may indicate a 3D anisotropic superconductor behavior [29–31]. We suppose a wider superconducting thickness leads to the phenomenon.

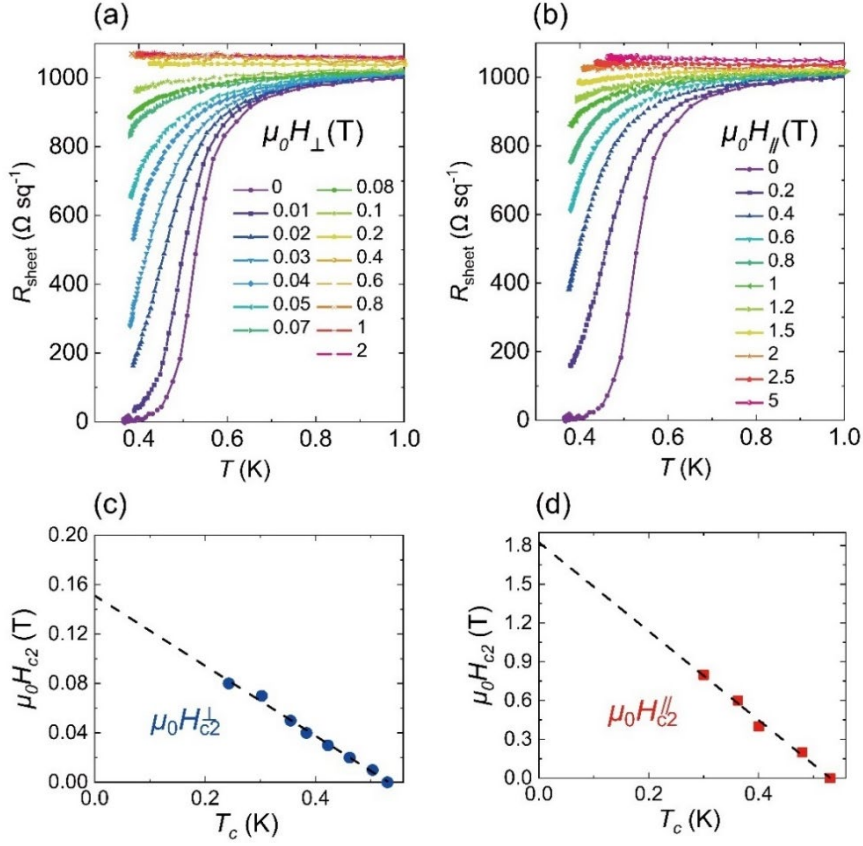

**Fig. 4** Transport behaviors under magnetic field for the LaVO$_3$(17 nm)/KTaO$_3$(111) heterostructure. Dependence of $R_{sheet}$ on $T$ for a field **a** perpendicular and **b** parallel to the interface. Temperature dependence of upper critical field $\mu_0 H_{c2}$, extracted from the 50% normal-state resistance, **c** perpendicular and **d** parallel to the interface.

**Oxygen vacancies and critical thickness**

As far as the interface conduction is concerned, we find that the oxygen pressure $P(O_2)$ during growth and the thickness of LaVO$_3$, $d_{LVO}$, are crucial. In Fig. 5a we display the $R_{sheet}(T)$ curves for LaVO$_3$/KTaO$_3$(111) heterostructures grown at different $P(O_2)$ values. The interfaces are highly insulating when the heterostructures were grown at $P(O_2) = 10^{-5}$ mbar and above, and metallic (superconducting) when grown at $P(O_2) = 10^{-6}$ mbar and below. This suggests strongly that oxygen vacancies play a determinant role in the interface conduction. Moreover, we find that while the heterostructure grown at $P(O_2) = 0$ mbar is superconducting, the one grown at $P(O_2) = 10^{-6}$ mbar is metallic but not superconducting. In addition, for the interfaces to be conducting, $d_{LVO}$ has to reach a critical thickness $d_c \sim 0.8$ nm. As shown in Fig. 5b, the sheet conductance $\sigma_s$ of the LaVO$_3$/KTaO$_3$(111) interface is below the measurement limit for $d_{LVO} < 0.8$ nm, and reaches saturation in an order of $10^{-4}$ $\Omega^{-1}$ for $d_{LVO} > 2$ nm. For a $d_{LVO}$ between 0.8 and 2 nm, the corresponding $\sigma_s$ can be large initially but decay severely with time in days (Fig. 5c).

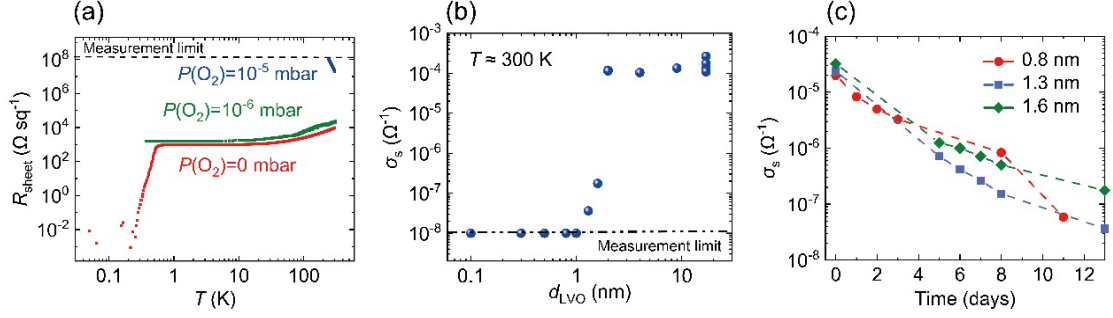

**Fig. 5** Effects of growth oxygen pressure and LaVO$_3$ thickness. **a** $R_{sheet}(T)$ curves for LaVO$_3$(17 nm)/KTaO$_3$(111) heterostructures grown under different oxgyen environments. **b** Room-temperature conductance $\sigma_s$ of LaVO$_3$/KTaO$_3$(111) heterostructures grown at $P(O_2) = 0$ mbar for different $d_{LVO}$ values. The data were measured after the conductance become stable. **c** Decay of $\sigma_s$ with time for LaVO$_3$/KTaO$_3$(111) heterostuctures of small $d_{LVO}$.

## 4 Discussion

First, we discuss the conduction mechanism of the epitaxial LaVO$_3$/KTaO$_3$ heterostructures. Previously we have demonstrated that the interface conduction in LaAlO$_3$/KTaO$_3$ heterostructures is controlled by electron transfer from oxygen vacancies in LaAlO$_3$ film to KTaO$_3$ substrate near the interface[19]. Here we propose that the same mechanism, that is, electron transfer from oxygen vacancies in LaVO$_3$ to KTaO$_3$, applies for the LaVO$_3$/KTaO$_3$ heterostructures. This scenario is supported by that a low $P(O_2)$, which can induce oxygen vacancies in LaVO$_3$, is necessary for the creation of interface conduction (Fig. 5a). It is further supported by that the interface conduction exhibits a gradual decay with time when $d_{LVO}$ is small, which can be explained by a gradual refilling of oxygen vacancies in LaVO$_3$ from ambient oxygen sources. Furthermore, note that although a low $P(O_2)$ can also induce oxygen vacancies in KTaO$_3$, we exclude them as a determinant conduction origin because even with them, the LaVO$_3$/KTaO$_3$ interface is still insulating if $d_{LVO}<d_c$ (including the bare KTO) (Fig. 5b). In LaAlO$_3$/SrTiO$_3$, after a long-standing debate[32–37], one widely accepted viewpoint[38] is that the key role of interface polar discontinuity is to form thermodynamically stable oxygen vacancies at the surface of LaAlO$_3$[38], which eventually transfer electrons to the interface. Therefore, the oxygen-vacancies-induced electron transfer and the classical electronic reconstruction share similar feature— electron transfer from oxygen vacancies in the film to the interface[38–40]. The key difference lies in the formation mechanism of oxygen vacancies. In LaVO$_3$/KTaO$_3$, it is reasonable to argue that the oxygen-deficient growth atmosphere, rather than the polar issue, is the main source of oxygen vacancies in LaVO$_3$.

Next, we discuss the superconductivity observed in the present epitaxial LaVO$_3$/KTaO$_3$(111) heterostructures. In the non-epitaxial EuO/KTaO$_3$ and LaAlO$_3$/KTaO$_3$ heterostructures, superconductivity was found at the (111) ($T_c$ ~2

K)[15,18] and (110) ($T_c$ ~0.9 K)[17] interfaces, but absent at the (001) interfaces[15,17]. A superconductivity of similar properties was also achieved on the bare KTaO$_3$ surfaces using an ionic liquid gating[28]. These results suggest that the KTaO$_3$ interface superconductivity is determined by the intrinsic properties of electron-doped surfaces of KTaO$_3$. Thus one would expect a superconductivity of similar properties in LaVO$_3$/KTaO$_3$ heterostructures. However, as shown in Fig. 3, only LaVO$_3$/KTaO$_3$(111) is superconducting, and its $T_c$ is much lower compared with that of the non-epitaxial ones (~0.5 K vs ~2 K). We speculate that the weaker superconductivity in LaVO$_3$/KTaO$_3$(111) is related to a wider conducting channel than that in the previous non-epitaxial ones.

## 5 Summary

In summary, we have fabricated epitaxial LaVO$_3$ film on KTaO$_3$ substrate and shown that a $T_c$ ~0.5 K superconductivity can be obtained in epitaxially grown LaVO$_3$/KTaO$_3$ (111) heterostructure. By contrast, no superconductivity is detected down to 50 mK in the epitaxially grown LaVO$_3$/KTaO$_3$(001) and LaVO$_3$/KTaO$_3$(110) heterostructures. Although in an earlier study[28], we have demonstrated that the presence of an oxide interface is not a prerequisite for the occurrence of KTaO$_3$ interface superconductivity, our present result, together with these previous ones[14–22], demonstrates that the detailed condition of an oxide interface still plays an important role in determining the superconductivity.

**Acknowledgements**

This work was supported by the National Natural Science Foundation of China (11934016, 12074334), the Key R&D Program of Zhejiang Province, China (2020C01019, 2021C01002), and the Fundamental Research Funds for the Central Universities of China.

**Author contributions**
Y.L., M.Z. and Y.S. fabricated and characterized the heterostructures. Z.L. and H.T. measured and analyzed the STEM data. Y.L. and Y.X. wrote the manuscript with input from all authors.

**Competing interests**
The authors declare no competing interests.

**Additional information**
**Supplementary Information** accompanies this paper.

# Supplementary Information for

Superconductivity in epitaxially grown $LaVO_3/KTaO_3(111)$

heterostructures

Yuan Liu[#], Zhongran Liu[#], Meng Zhang, Yanqiu Sun, He Tian, and Yanwu Xie*

#These authors contributed equally to this work.
Correspondence: ywxie@zju.edu.cn

**This file includes:**

    Figs. S1-S6

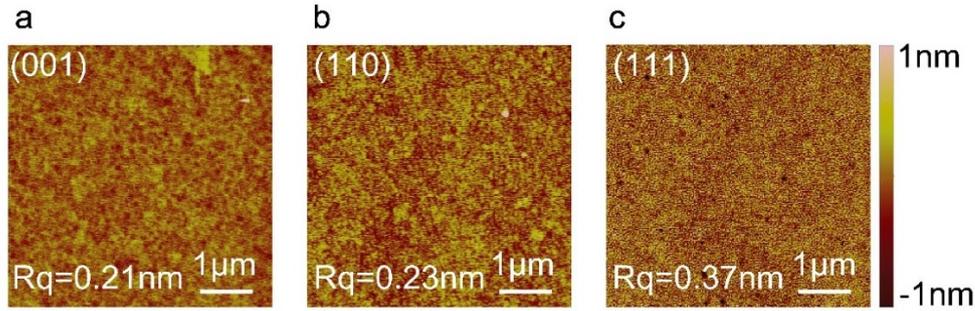

**Fig. S1** Atomic force microscopy images of 17-nm LaVO$_3$ thin films grown on KTaO$_3$ substrates with different out-of-plane crystalline orientations. All these surfaces are smooth. The measured root mean square roughness (over 5 × 5 μm$^2$) is **a** 0.21 nm, **b** 0.23 nm, and **c** 0.37 nm for (001), (110), and (111), respectively.

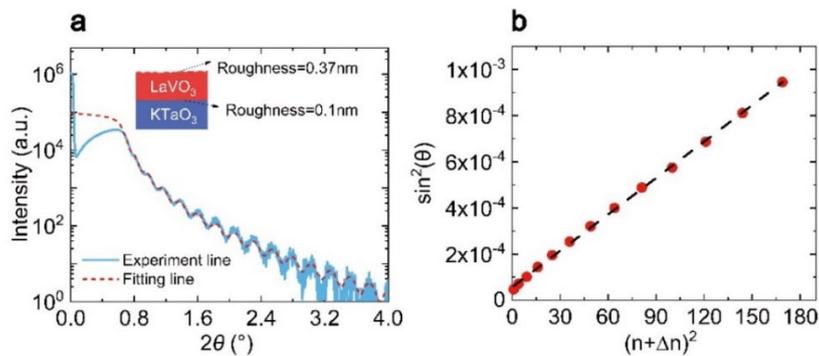

**Fig. S2** A Small-angle x-ray reflectivity measurement of a LaVO$_3$/KTaO$_3$(111) heterostructure, red dashed line is fitting to the data. **b** Fitting of the oscillations in **a** by $(\sin\theta)^2 \propto (n+\Delta n)^2 (\lambda/2d)^2$ give a thickness $d$ of ~35 nm. Here $\lambda$ is the wavelength of Cu $K\alpha_1$ radiation

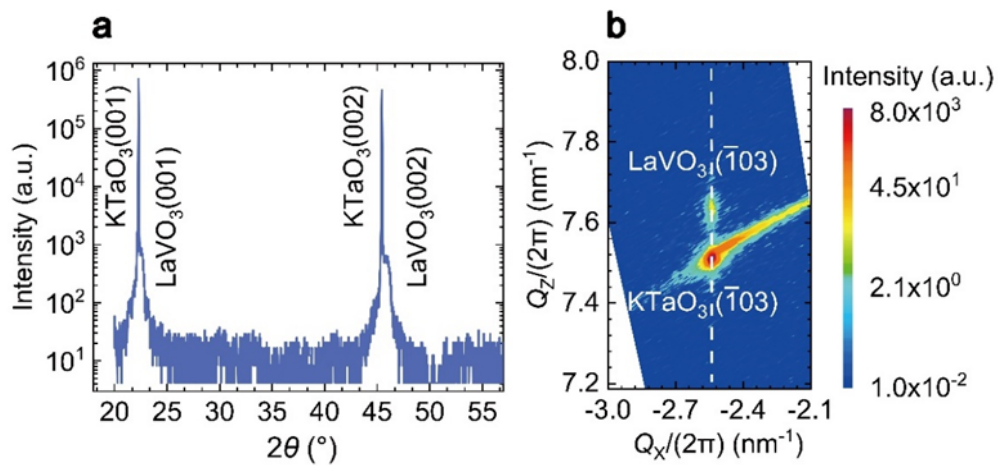

**Fig. S3** X-ray diffraction of a LaVO$_3$(17 nm)/KTaO$_3$(001) heterostructure. **a** Out-of-plane $\theta$-$2\theta$ XRD pattern. **b** Reciprocal space mapping around the (-103) reflection. LaVO$_3$ and KTaO$_3$ Bragg's peaks aligned along the vertical dashed line.

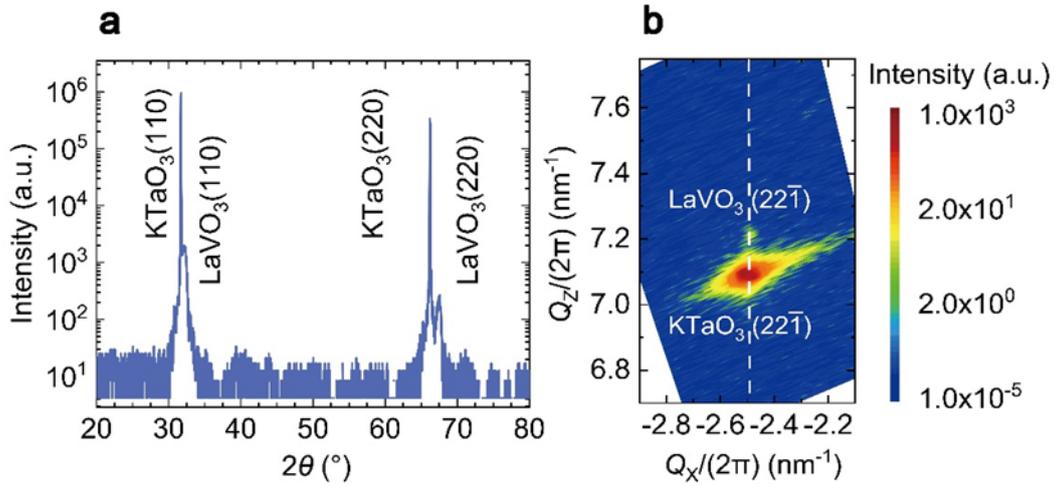

**Fig. S4** X-ray diffraction of a LaVO$_3$(17 nm)/KTaO$_3$(110) heterostructure. **a** Out-of-plane $\theta$-$2\theta$ XRD pattern. **b** Reciprocal space mapping around the (22-1) reflection. LaVO$_3$ and KTaO$_3$ Bragg's peaks aligned along the vertical dashed line.

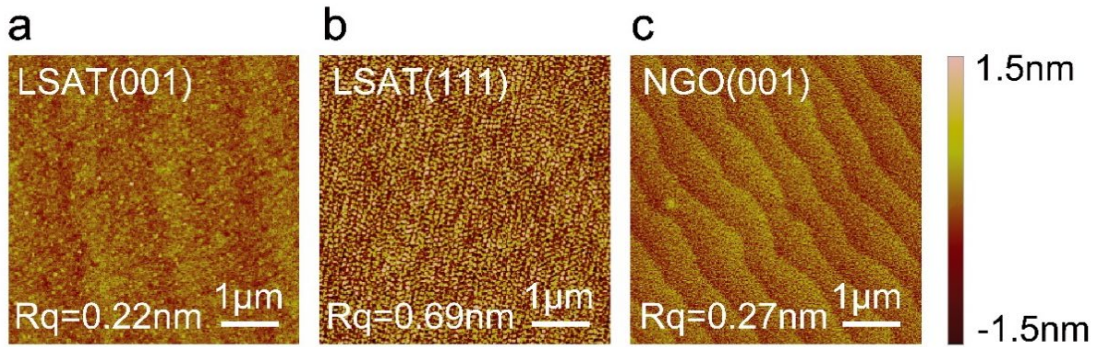

**Fig. S5** Atomic force microscopy images of 17-nm LaVO$_3$ thin films grown on **a** (La$_{0.3}$Sr$_{0.7}$)(Al$_{0.65}$Ta$_{0.35}$)O$_3$ (LSAT) (001), **b** LSAT(111), and **c** NdGaO$_3$(NGO)(001) single-crystalline substrates. The root mean square roughness of the each surface (over 5 × 5 μm$^2$) is as labelled. These control samples were grown under conditions identical to the standard ones described in the main text.

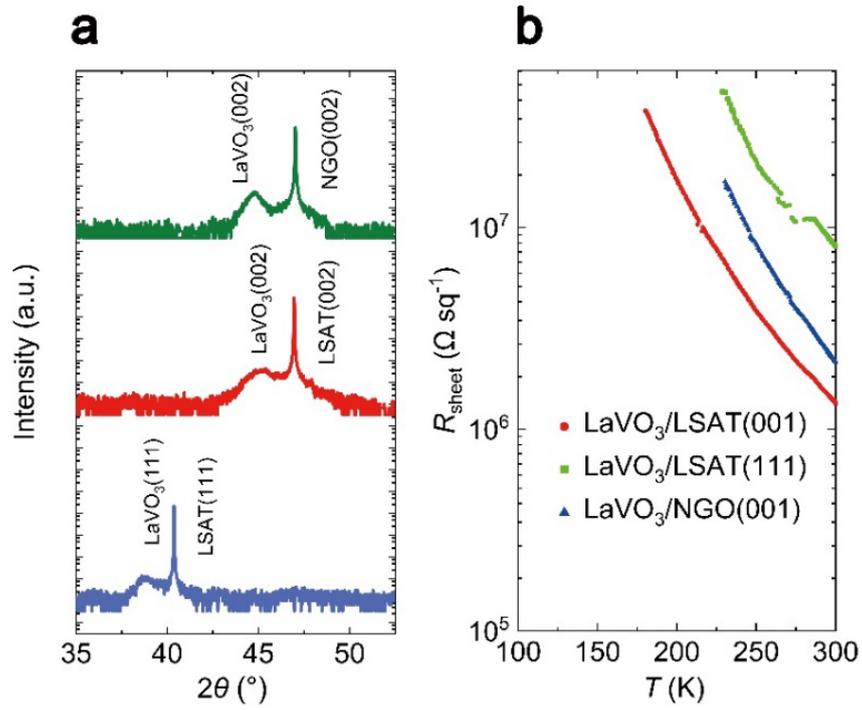

**Fig. S6** Structure and transport measurements on control samples (the same ones as in Fig. S5). **a** Out-of-plane $\theta$-$2\theta$ XRD patterns. **b** Temperature-dependent $R_{sheet}$ curves. These data show that the epitaxially grown $LaVO_3$ film itself is insulating, and its $R_{sheet}$ values in a temperature range from 300 K to low temperatures are extremely large.